\documentclass[11pt]{article}	
\usepackage{psfig}
\begin{document}
\pagenumbering{arabic}
\newcommand{\ket}[1]{\left | #1 \right \rangle}

\title{A Quantum Algorithm for finding the Maximum}

%\author{  Ashish Ahuja  \{ashish@cse.iitd.ernet.in\}\\
%          Sanjiv Kapoor  \{skapoor@cse.iitd.ernet.in\}

\date{}
\author{  Ashish Ahuja $~~~~~~~~~~$  Sanjiv Kapoor\thanks{email : skapoor@cse.iitd.ernet.in}\\Department of Computer Science and Engineering\\
Indian Institute of Technology, New Delhi}

\maketitle

\begin{abstract}
{\it 
This paper describes a quantum algorithm for finding the maximum among
$N$ items. The classical method for the same problem takes $O(N)$
steps because we need to compare two numbers in one step. 
This algorithm takes $O(\sqrt{N})$ steps by exploiting the property of quantum
states to exist in a superposition of states and hence performing
an operation on a number of elements in one go. A tight upper
bound of $6.8\sqrt{N}$ for the number of steps needed using this algorithm was found. These steps are the number of queries made to the oracle.
}
\end{abstract}

\section{Introduction}
Quantum computing as a new paradigm of computing was introduced in the early 1980's in the work of Feynman~\cite{Feynman} and Beinoff~\cite{Beinoff}. Later they were shown to be atleast as powerful as classical computers. But the real interest in exploring the bridge between physics and computation arose when quantum algorithms which improved over their classical counterparts were proposed. One was the algorithm for factorization in polynomial number of steps by Shor~\cite{Shor} and another was algorithm which achieved quadratic speedup for the classic problem of database search by Grover~\cite{Grover}. This power of quantum computation is due to the fact that     
the state of a quantum computer can be a superposition of basis states and
we can perform a operation on multiple quantum states simultaneously. 
This paper applies the quantum techniques to a very basic problem in
order statistics i.e. of finding the extrema of a set of
numbers.

\section{Problem} Let $T\lbrack 0\ldots N-1\rbrack$ be an unsorted table of $N$
distinct items. We have to find the index $i$ of the element such that $T\lbrack i\rbrack$ is maximum.

\section{Algorithm}

The algorithm below uses an oracle that computes:

\[ f_i(j) = \left \{  \begin{array}{ll}
  	1 &  \mbox{if $T\lbrack j \rbrack > T\lbrack i \rbrack$} \\
	0 & \mbox{otherwise}\\
	\end{array}
\right. \]
		This algorithm starts out with an initial guess of the
	index of the maximum and using Grover's search algorithm, finds
	out a new element which is the index of one of the elements
	greater than the initial indexed element. The new index forms
	our new guess and the whole procedure is repeated. The various
	steps of the algorithm are : 

\begin{enumerate}

\item Choose a index $y$ randomly from the set $\lbrace 0..N-1
\rbrace$. This forms our initial guess of the index of the maximum element.

\item Repeat the following $O(\sqrt{N})$ times:

\begin{itemize}

\item Initialize the state as $\ket{\psi} =
\sum_i{\frac{1}{\sqrt{N}}\ket{i}\ket{y}}$. This can be achieved in $O(logN)$ steps using a $n$-bit Hadamard transformation. 

\item Apply the Grover's search algorithm for finding a  marked
state (if no. of marked states is $\geq$ 1)~\cite{Grover}. We consider a state $x$ as
marked if the following is satisfied,
\[f_y(x)= 1.\]
\item Make a measurement on first register. Let the result of the
	measurement be $x_0$. Then our new guess $y$ will be $x_0$.
\end{itemize}

\item Return $y$ as the index of the maximum element in the table.

\end{enumerate}

\section{Complexity Analysis} Let $E(N,t)$ be the expectation value of
the number of iterations for finding the maximum of $N$ items in which
t are marked. Then we write a recurrence equation for $E(N,t)$ as :

\begin{equation}
E(N,t) = \frac{1}{t}\lbrack E(N,t-1)+....E(N,1) \rbrack +
6\sqrt{\frac{N}{t}}
\end{equation}
(Since the expected no. of iterations to find the index of a marked
item among $N$ items where $t$ are marked is at most $6\sqrt{N/t}$)~\cite{Brassard}.
Writing the equations when the no. of marked items is $t$ and ($t-1$), we
get
\begin{equation}
E(N,t) = \frac{1}{t}\sum_{i=1}^{t-1}E(N,i) + 6\sqrt{\frac{N}{t}}
\end{equation}

\begin{equation}
E(N,t-1) = \frac{1}{t-1}\sum_{i=1}^{t-2}E(N,i) + 6\sqrt{\frac{N}{t-1}}
\end{equation}
\\
Multiplying (2) by $t$ and (3) by ($t-1$) we get,
\begin{equation}
tE(N,t) = \sum_{i=1}^{t-1}E(N,i) + 6\sqrt{Nt}
\end{equation}

\begin{equation}
(t-1)E(N,t-1) = \sum_{i=1}^{t-2}E(N,i) + 6\sqrt{N(t-1)}
\end{equation}
\\
Subtracting (5) from (4) and rearranging, we get
\begin{equation}
E(N,t) = E(N,t-1) + 6\frac{\sqrt{N}}{t}(\sqrt{t}-\sqrt{t-1})
\end{equation}
\\
Writing the same equation for ($t-1$),...,2 and adding all of them, we get,
\begin{equation}
E(N,t) = E(N,1) + 6\sqrt{N}\sum_{i=2}^{t}\frac{1}{i}(\sqrt{i}-\sqrt{i-1})
\end{equation}
This can be rewritten as :
\begin{equation}
E(N,t) = E(N,1) + 6\sqrt{N}\sum_{i=2}^{t}\frac{1}{\sqrt{i}}(1-\sqrt{1-\frac{1}{i}})
\end{equation}
Applying the relation $(1-x)^n > (1-nx)$ and replacing summation by
integral, we get
\begin{equation}
E(N,t) \leq E(N,1) + 6\sqrt{N}\int_{i=1}^{t}\frac{1}{2z\sqrt{z}}dz.
\end{equation}
Evaluating the integral,we get 
\begin{equation}
E(N,t) \leq E(N,1) + 6\sqrt{N}(1-\frac{1}{\sqrt{t}}).
\end{equation}
An upper bound is achieved for $t = N-1$, when we get
\[E(N,t)\leq 6.8\sqrt{N}\]
		Thus,we get the upper bound on the number of
		iterations as $6.8\sqrt{N}$. After we had the initial idea of this algorithm, Dr.Grover told us that there already exists a minimum searching algorithm by D{\"u}rr and H\o yer~\cite{Hoyer}. We found that both the algorithms were the same. Our complexity analysis yields a better bound than the original one of $15\sqrt{N}$. We can repeat this algorithm to obtain the maximum with a probability of $1-(\frac{1}{2})^k$ after $k$ repetitions of this algorithm.

	We next show a high probability bound for the problem. Applying Markov's inequality, we have, 
	\[ P(X \geq k\langle X \rangle) \leq \frac{1}{k} \]
	where $\langle X \rangle$ is the expected value of the random variable $X$. Therefore using the upper bound on the expected number of iterations, we get 
 	\[ P(X \geq 6.8k\sqrt{N}) \leq \frac{1}{k} \]
	So after  $13.6\sqrt{N}$ steps, we would have reduced the probability of failure to less than $1/2$. Repeated iterations of this algorithm give us a high probability bound but with an additional factor of $O(logN)$. It would be of interest to improve this bound further.

\section*{Acknowledgments}

We are thankful to Lov Grover and Peter H{\o}yer for going through the initial manuscript
and pointing out some errors.

\end{document}